\documentclass[letterpaper,english,preprint]{revtex4}
\usepackage[T1]{fontenc}
\usepackage[latin1]{inputenc}

\makeatletter

\providecommand{\tabularnewline}{\\}

\usepackage{array}
\usepackage{amsmath}
\usepackage{color}
\usepackage{graphicx}
\usepackage{amssymb}
\usepackage{graphicx}

\usepackage{babel}

\usepackage{babel}
\makeatother
\begin{document}

\title{THE LANDAU DISTRIBUTION FOR CHARGED PARTICLES TRAVERSING THIN FILMS}

\author{M. MARUCHO}

\address{Department of Chemistry, Univeristy of Houston,\\
 Houston, Texas, 77204-5003, USA.\\
mmarucho@uh.edu}

\author{C.A. GARCIA CANAL$^{*}$ AND HUNER FANCHIOTTI$^{**}$}

\address{Laboratorio de F\'{\i}sica Te\'{o}rica, Departamento de F\'{\i}sica,
Universidad Nacional de La Plata,\\
C.C. 67 - 1900 La Plata, Argentina.\\
$^{*}$garcia@fisica.unlp.edu.ar, $^{**}$huner@fisica.unlp.edu.ar}

\begin{abstract}
The Landau distribution as well as its first and second momenta are
well suited for describing the energy loss of charged particles traversing
a thin layer of matter. At present, just rational approximations and
asymptotic expressions for these functions were obtained. In this
paper we present a direct calculation of the integral representation
of these functions obtaining perturbative and nonperturvative solutions
expressed in terms of fast convergent series. We also provide a simple
numerical algorithm which allows to control speed and precision of
the results. The testing runs have provided, in reasonable computing
times, correct results up to 13-14 significant digits on the density
and distribution functions and 9-10 on the first and second momenta.
If necessary, this accuracy could be improved by adding more coefficients
to the algorithm.


\end{abstract}
\maketitle

\section{Introduction}

It is well known that the Bethe-Bloch formula\cite{bethe} describes
the average energy loss of charged particles when travelling through
matter, while the fluctuations of energy loss by ionization of a charged
particle in a thin layer of matter was theoretically described by
Landau\cite{landau}. This description ends with a universal asymmetric
probability density function characterized by a narrow peak with a
long tail for positive values. This tail towards positive values comes
from the small number of individual collisions, each with a small
probability of transferring comparatively large amounts of energy.
An integral representation of the Landau probability density function
reads

\begin{equation}
\phi\left(x\right)=\frac{1}{\pi}\int_{0}^{\infty}y^{-y}\sin\left(\pi y\right)\exp\left(-xy\right)dy\label{eqlandau}\end{equation}
 with $x>0$.

Other functions such as the corresponding distribution function $\Phi\left(0,x\right)$
and the first and second momenta $\Phi\left(1,x\right)$ and $\Phi\left(2,x\right)$
of the density function truncated on the right-hand tail defined through
the general formula

\begin{equation}
\Phi\left(n,x\right)\left[\frac{\left(n-1\right)\left(n-2\right)}{2}+\frac{n\left(3-n\right)}{2}\Phi\left(0,x\right)\right]=\int_{-\infty}^{x}c^{n}\phi\left(c\right)dc\label{eq:moments}\end{equation}
 are also needed in order to fit a truncated Landau distribution $\Phi\left(0,x\right)$
to measurements or to simulated energy-loss data\cite{landau,selt}.

It is clear that a direct calculation, by numerical integration, of
expressions (\ref{eqlandau}) and (\ref{eq:moments}) besides of being
of a very slow convergence could be used only to provide values for
a subsequent construction of approximate formula and for testing purposes.

In the seventies some programs for the numerical computation of the
Landau distribution were presented\cite{schorr1}. There, the main
emphasis was put more on the speed of the calculation algorithms than
in the precision obtained, that was in general not good enough. In
the eighties a new package of programs\cite{schorr2} arouse. These
programs improved the precision of the values obtained for (\ref{eqlandau})
and (\ref{eq:moments}) by using piecewise functions as rational approximations
and considering their asymptotic behavior. One should mention also
Ref. 6 were more accurate calculations of the Landau density were
presented. For more recent articles see Ref. 7. In view of the still
high computational time required, the precision for the distribution
function was only up to 6-7 significant digits. At that time, this
was thought to be enough for the applications in experimental physics
but their use in other fields of application remained excluded.

For practical uses and under the standard prejudice that equation
(\ref{eqlandau}) cannot be rewritten into an analytic form, the following
approximate expression was suggested\cite{moyal}

\begin{equation}
\phi\left(x\right)\simeq\exp\left[-\frac{1}{2}\left(x+\exp\left(-x\right)\right)\right]/\sqrt{2\pi}.\label{eqmoyal}\end{equation}

This analytic approximation, known as the Moyal function, is in fact
an extreme value distribution\cite{kotz}. It provides values that
are not entirely coincident with those coming from the original expression
(\ref{eqlandau}), in particular in connection with its maximum value
and its asymptotic behavior. Nevertheless, being quite simple, it
is widely used in different fields. See Ref. 10 for examples.

To overcome the aforementioned difficulties, new and refined approximations,
combining speed and accuracy in the evaluation of these functions,
are required to improve the agreement between measured and calculated
energy loss distributions.

Under this motivation, we have carried out a direct calculation of
the integral representation of the Landau density, the distribution
function and the two first momenta of the density. We have obtained
perturbative and nonperturvative solutions expressed in terms of fast
convergent series. To achieve this, we have used analytical continuation
properties and the Laplace transform of the Landau density. We have
also developed a simple numerical algorithm which allows one to control
speed and precision of the results.

In what follows we summarize the main steps for obtaining the explicit
analytic expressions for the Landau probability density (\ref{eqlandau})
and the momenta (\ref{eq:moments}). An appendix with details of the
calculation is annexed. In addition, a brief description of the algorithm
which evaluates numerically these functions is presented. Finally,
a numerical test together with a critical analysis of the results
is provided.

\section{Nonperturbative Solutions}

\subsection{Landau density}

For computational purposes, it is of interest to consider the following
auxiliary integral

\begin{equation}
Q\left(x,b\right)=\frac{1}{\pi}\int_{0}^{\infty}y^{-y}\exp\left[-y\left(x+b\right)\right]dy.\end{equation}

Clearly, once its solution is known, it is necessary to perform the
analytic continuation

\begin{equation}
\phi\left(x\right)={\textrm{Im}}\left[Q\left(x,b=-i\pi\right)\right]\label{anacon}\end{equation}
 to obtain the Landau density.

By using standard representations of the functions in the integrand
of $Q(c,b)$, one can write

\[
Q\left(x,b\right)=\frac{1}{\pi}\sum_{k=0}^{\infty}\frac{\left(-\right)^{k}}{k!}\frac{\partial^{k}}{\partial a^{k}}\left[\frac{\Gamma\left(k+a+1\right)}{\left(x+b\right)^{k+a+1}}\right]_{a=0}\]
 where $\Gamma$ is the gamma function\cite{abra}. Using now the
identity \[
\frac{\partial^{k}}{\partial a^{k}}\left[F\left(a\right)G\left(a\right)\right]=\sum_{r=0}^{k}\left(\begin{array}{c}
k\\
r\end{array}\right)\frac{\partial^{k-r}}{\partial a^{k-r}}\left[F\left(a\right)\right]\frac{\partial^{r}}{\partial a^{r}}\left[G\left(a\right)\right]\]
 and defining the coefficients \begin{equation}
A_{kr}\equiv\left[\frac{\partial^{k-r}}{\partial a^{k-r}}\Gamma\left(k+a+1\right)\right]_{a=0}\label{eq:akr}\end{equation}
 one obtains

\[
Q\left(x,b\right)=\frac{1}{\pi}\sum_{k=0}^{\infty}\frac{\left(-\right)^{k}}{k!}\sum_{r=0}^{k}\left(\begin{array}{c}
k\\
r\end{array}\right)\left[\frac{A_{kr}}{\left(x+b\right)^{k+1}}\right]\left(-\right)^{r}\left[\ln\left(x+b\right)\right]^{r}.\]
 Finally, performing the analytic continuation (\ref{anacon}), one
obtains the exact analytic expression for the Landau density

\begin{equation}
\phi\left(x\right)=\frac{1}{\pi}\sum_{k=0}^{\infty}\frac{\left(-\right)^{k}}{k!}\sum_{r=0}^{k}\left(\begin{array}{c}
k\\
r\end{array}\right)\left(-\right)^{r}A_{kr}{\textrm{Im}}\left[\frac{\left[\ln\left(x-i\pi\right)\right]^{r}}{\left(x-i\pi\right)^{k+1}}\right].\label{eq:density}\end{equation}

Notice that this expression can be analytically continued to $x\leq0$
values. However, we obtain below a formal solution also for this case,
since the numerical evaluation of eq.(\ref{eq:density}) for negative
values of $x$ makes the series divergent.

\subsection{Landau distribution and momenta}

In order to compute the functions defined in Eq.(\ref{eq:moments})
we use the Laplace transform of the Landau density, namely \[
\int_{-\infty}^{+\infty}\exp\left(-pc\right)\phi\left(c\right)dc=p^{p}\quad;\quad p\geq0.\]

The derivative of this relation with respect to $p$, gives rise to
a new and convenient expression for computing the Landau distribution
and the two first momenta of the density as follows

\begin{equation}
\begin{array}{c}
{\displaystyle \Phi\left(n,x\right)\left[\frac{\left(n-1\right)\left(n-2\right)}{2}+\frac{n\left(3-n\right)}{2}\Phi\left(0,x\right)\right]=}\\
{\displaystyle \begin{array}{l}
\lim\\
p\rightarrow0^{+}\end{array}\left[\left(-\right)^{n}\left[\frac{n\left(n-1\right)\left(n-3/2\right)}{p}+\left(1+\ln\left(p\right)\right)^{n}\right]p^{p}\right.}\\
{\displaystyle \left.-\int_{x}^{+\infty}c^{n}\, e^{-pc}\,\phi\left(c\right)dc\right].}\end{array}\label{eq:gralexpression}\end{equation}

Replacing now the Landau density (\ref{eq:density}), into this last
expression, we see that the calculation is reduced to the computation
of

\begin{equation}
\begin{array}{c}
{\displaystyle G\left(p,x,n\right)\equiv\frac{1}{\pi}\sum_{k=0}^{\infty}\frac{\left(-\right)^{k}}{k!}\sum_{r=0}^{k}\left(\begin{array}{c}
k\\
r\end{array}\right)\left(-\right)^{r}A_{kr}}\\
{\displaystyle \begin{array}{l}
\lim\\
p\rightarrow0^{+}\end{array}{\textrm{Im}}\left[\int_{x}^{+\infty}c^{n}\, e^{-pc}\frac{\left[\ln\left(c-i\pi\right)\right]^{r}}{\left(c-i\pi\right)^{k+1}}dc\right]}\end{array}\label{eq:q}\end{equation}
 that means the integrals \begin{equation}
B_{kr}^{n}\left(p,x\right)=\begin{array}{l}
\lim\\
p\rightarrow0^{+}\end{array}{\textrm{Im}}\left[\int_{x}^{+\infty}c^{n}\, e^{-pc}\frac{\left[\ln\left(c-i\pi\right)\right]^{r}}{\left(c-i\pi\right)^{k+1}}dc\right].\label{eq:integrals}\end{equation}

Once these integrals are evaluated, an analysis of their eventual
divergences and the necessary cancellations are in order.

\subsubsection{Landau distribution}

Let us solve first the case $n=0$, the Landau distribution function.
Notice that in this case all the integrals in expression (\ref{eq:integrals})
are well defined and consequently the interchange between the limit
operation and the integral is licit. In this way we obtain

\begin{equation}
B_{00}^{0}\left(0,x\right)={\textrm{Im}}\left[\int_{x}^{+\infty}\frac{1}{\left(c-i\pi\right)}dc\right]=\pi\left[1/2-\arctan\left(x/\pi\right)/\pi\right]\label{eq:nzerokzero}\end{equation}
 and

\begin{equation}
B_{kr}^{0}\left(0,x\right)=\begin{array}{l}
\lim\\
b\rightarrow\infty^{+}\end{array}{\textrm{Im}}\left[\int_{\ln\left(x-i\pi\right)}^{\ln\left(b-i\pi\right)}u^{r}\exp\left(-ku\right)du\right].\label{eq:nzeroknonzero}\end{equation}
 A further change of variables to $v=uk$, reduces the integral to
an incomplete Gamma function\cite{abra}. Finally, after taking the
limit, one obtains

\[
B_{kr}^{0}\left(0,x\right)=\frac{1}{k^{r+1}}{\textrm{Im}}\left[\Gamma\left(r+1,k\ln\left(x-i\pi\right)\right)\right].\]
 Going back to equation (\ref{eq:q}) one gets

\[
G\left(0,x,0\right)=\frac{1}{2}-\frac{1}{\pi}\,\arctan\left(\frac{x}{\pi}\right)+\frac{1}{\pi}\sum_{k=1}^{\infty}\frac{\left(-\right)^{k}}{k!}\sum_{r=0}^{k}\left(\begin{array}{c}
k\\
r\end{array}\right)\left(-\right)^{r}A_{kr}B_{kr}^{0}\left(0,x\right)\]
 so that the Landau distribution results

\begin{equation}
\begin{array}{c}
{\displaystyle \Phi\left(0,x\right)=\frac{1}{2}+\frac{1}{\pi}\,\arctan\left(\frac{x}{\pi}\right)}\\
{\displaystyle -\frac{1}{\pi}\sum_{k=1}^{\infty}\frac{\left(-\right)^{k}}{k!}\sum_{r=0}^{k}\left(\begin{array}{c}
k\\
r\end{array}\right)\frac{\left(-\right)^{r}A_{kr}}{k^{r+1}}{\textrm{Im}}\left[\Gamma\left(r+1,k\ln\left(x-i\pi\right)\right)\right].}\end{array}\label{eq:distribution}\end{equation}

\subsubsection{Momenta}

Going now to the analysis of the momenta, we found that in the general
expression there appear divergences that finally cancel out, but that
have to be treated with care. We postpone to an Appendix some details
of the lengthy calculation involved and present here the final results,
both for the first and the second momenta. They read

\[
\Phi\left(1,x\right)\Phi\left(0,x\right)=-1+\gamma+\frac{1}{2}\ln\left(x^{2}+\pi^{2}\right)\]

\[
+\frac{1}{\pi}\sum_{r=0}^{1}\left(-\right)^{r}A_{1r}\left[\pi Re\left[\Gamma\left(r+1,\ln\left(x-i\pi\right)\right)\right]+\right.\]

\[
\left.\frac{1}{r+1}\left[1-r+r\ln\left(x^{2}+\pi^{2}\right)\right]\left[\pi/2-\arctan\left(x/\pi\right)\right]\right]-\frac{1}{\pi}{\textrm{Im}}\sum_{m=0}^{1}\left(i\pi\right)^{1-m}\]

\begin{equation}
\sum_{k=2}^{\infty}\frac{\left(-\right)^{k}}{k!}\sum_{r=0}^{k}\left(\begin{array}{c}
k\\
r\end{array}\right)\frac{\left(-\right)^{r}A_{kr}}{\left(k-m\right)^{r+1}}\left[\Gamma\left(r+1,\left(k-m\right)\ln\left(x-i\pi\right)\right)\right]\label{eq:momentouno}\end{equation}

\noindent and

\[
\Phi\left(2,x\right)\Phi\left(0,x\right)=1+x+\frac{\pi}{2}\left[\pi-2\arctan\left(x/\pi\right)\right]\]

\[
+\left(1-\gamma\right)\left[\frac{x^{2}}{\left(x^{2}+\pi^{2}\right)}-1-2\left[\gamma+\frac{1}{2}\ln\left(x^{2}+\pi^{2}\right)\right]\right]\]

\[
-\frac{1}{\pi}\left[\frac{x^{2}\left[x\left[\arctan\left(x/\pi\right)-\frac{\pi}{2}\right]+\frac{\pi}{2}\ln\left(x^{2}+\pi^{2}\right)+\pi\right]}{x^{2}+\pi^{2}}-2\pi+\gamma\pi+\gamma^{2}\pi+\frac{\pi^{3}}{6}\right]\]

\[
-\frac{1}{\pi}\left[-\frac{\pi}{4}\left[\ln\left(x^{2}+\pi^{2}\right)\right]^{2}-2x\left[\arctan\left(x/\pi\right)-\frac{\pi}{2}\right]+\pi\left[\arctan\left(x/\pi\right)-\frac{\pi}{2}\right]^{2}\right]\]

\[
-\frac{1}{2\pi}\sum_{r=0}^{2}\left(\begin{array}{c}
2\\
r\end{array}\right)\left(-\right)^{r}A_{2r}{\textrm{Im}}\left[-\frac{\left[\ln\left(x-i\pi\right)\right]^{r+1}}{r+1}\right.\]

\[
\left.+\sum_{m=0}^{1}\left(\begin{array}{c}
2\\
m\end{array}\right)\frac{\left(i\pi\right)^{2-m}}{\left(2-m\right)^{r+1}}\left[\Gamma\left(r+1,\left(2-m\right)\ln\left(x-i\pi\right)\right)\right]\right]\]

\[
-\frac{1}{\pi}{\textrm{Im}}\sum_{m=0}^{2}\left(\begin{array}{c}
2\\
m\end{array}\right)\left(i\pi\right)^{2-m}\sum_{k=3}^{\infty}\frac{\left(-\right)^{k}}{k!}\]

\begin{equation}
\sum_{r=0}^{k}\left(\begin{array}{c}
k\\
r\end{array}\right)\frac{\left(-\right)^{r}A_{kr}}{\left(k-m\right)^{r+1}}\left[\Gamma\left(r+1,\left(k-m\right)\ln\left(x-i\pi\right)\right)\right]\label{eq:momentodos}\end{equation}
 being $\gamma$ the Euler's constant\cite{abra}. Notice that the
limit $x\rightarrow\infty$ indicates that these distributions, even
if normalized to $1$, do not have well defined momenta.

\section{Perturbative Solutions}

\subsection{Landau density}

For the sole purpose of presenting solutions in an as simple as possible
way, we define two auxiliary functions $M_{j}\left(x\right)$ and
$T_{ijl}\left(x\right)$. First

\begin{equation}
M_{j}\left(x\right)\equiv\frac{1}{\pi}\sum_{n=0}^{\infty}\frac{\left(-x\right)^{n}}{n!}\sum_{k=1}^{\infty}\left(-\right)^{k}\exp\left(-xk\right)\sum_{p=0}^{\infty}\frac{D_{jnkp}}{\left(2p+1\right)!4^{p}},\label{aux1}\end{equation}
 where the coefficients $D_{jnkp}$ are given by

\begin{equation}
D_{jnkp}=\left.\frac{\partial^{2p}}{\partial u^{2p}}\left(u+k\right)^{\left(-j-1-u-k\right)}u^{n}\sin\left(\pi u\right)\right|_{u=1/2}.\label{coef2}\end{equation}

Note that the expression (\ref{aux1}) is the pertubative solution
of the integral

\[
M_{j}\left(x\right)=\frac{1}{\pi}\int_{1}^{\infty}y^{-y-j-1}\sin\left(\pi y\right)\exp\left(-xy\right)dy.\]

This solution is simply obtained by dividing the integration range
in consecutive unit segments and expanding afterwards the exponential
and trigonometric functions in the integrand.

The other auxiliary function is given by

\begin{equation}
T_{ijl}\left(x\right)\equiv\frac{\left(-1\right)^{j}}{\pi}\sum_{n=0}^{\infty}\frac{\left(-x\right)^{n}}{n!}t_{ijln}\label{aux2}\end{equation}
 being the coefficients $t_{ijln}$ given by the following expression

\[
t_{ijln}=\sum_{m=0}^{\infty}\frac{\left(-1\right)^{m}\pi^{2m+l}}{\left(2m+l\right)!}\left[1-\frac{l\left(l-1\right)/2}{2m+3}\right]\sum_{k=0}^{\infty}\frac{\left(k+j\right)!}{k!\left(k+n+i+2m+l+1\right)^{k+j+1}}.\]

The expression (\ref{aux2}) is the perturbative solution of the integrals

\[
\frac{1}{\pi}\int_{0}^{1}y^{-y+i}\cos\left(\pi y\right)\exp\left(-xy\right)\ln^{j}\left(y\right)dy,\frac{1}{\pi}\int_{0}^{1}y^{-y+i}\sin\left(\pi y\right)\exp\left(-xy\right)\ln^{j}\left(y\right)dy\]
 and

\[
\frac{1}{\pi}\int_{0}^{1}y^{-y+i}\left[-\cos\left(\pi y\right)+\frac{\sin\left(\pi y\right)}{\pi y}\right]\exp\left(-xy\right)\ln^{j}\left(y\right)dy\]
 for $l=0,1$ and $2$ respectively, which is obtained by expanding
the exponential $\exp\left(-y\ln\left(y\right)\right),$ and the other
functions appearing in the integrand, in powers of their arguments.

As a consequence, the Landau density is obtained by dividing the range
of integration in expression (\ref{eqlandau}) into $\left[0,1\right]\times\left[1,\infty\right]$.
Thus, we get

\begin{equation}
\phi\left(x\right)=T_{001}\left(x\right)+M_{-1}\left(x\right).\label{densitypert}\end{equation}

\subsection{Landau distribution and momenta}

The computation of these functions follows the steps done for the
case of the nonperturbative regime. In this case, the integrals appearing
in Eq.(\ref{eq:gralexpression}) are now computed from the following
relationship

\begin{equation}
\int_{x}^{+\infty}c^{n}\, e^{-pc}\,\phi\left(c\right)dc=\frac{1}{\pi}\sum_{j=0}^{n}\left(\begin{array}{c}
n\\
j\end{array}\right)j!x^{n-j}\int_{0}^{\infty}\frac{y^{-y}}{\left(p+y\right)^{j+1}}\sin\left(\pi y\right)\exp\left(-xy\right)dy.\label{gral2}\end{equation}

In the limit $p$ going to zero, Eq.(\ref{eq:gralexpression}) evaluated
at $n=0$ becomes

\begin{equation}
\Phi\left(0,x\right)=1-\frac{1}{\pi}\int_{0}^{\infty}\frac{y^{-y}}{\left(p+y\right)}\sin\left(\pi y\right)\exp\left(-xy\right)dy.\label{distrib2}\end{equation}

Let us observe that this integral is well defined allowing us to evaluate
it at $p=0$. In this way, we obtain the following expression for
the Landau Distribution

\begin{equation}
\Phi\left(0,x\right)=1-\frac{1}{\pi}\int_{0}^{\infty}y^{-y-1}\sin\left(\pi y\right)\exp\left(-xy\right)dy.\label{distrib3}\end{equation}

Analogously, we get for $n=1$ and $n=2$

\begin{equation}
\Phi\left(1,x\right)\Phi\left(0,x\right)=-1-\ln\left(p\right)-\frac{x}{\pi}\int_{0}^{\infty}y^{-y-1}\sin\left(\pi y\right)\exp\left(-xy\right)dy\label{m1}\end{equation}

\[
-\frac{1}{\pi}\int_{0}^{\infty}\frac{y^{-y}}{\left(p+y\right)^{2}}\sin\left(\pi y\right)\exp\left(-xy\right)dy\]
 and

\[
\Phi\left(2,x\right)\Phi\left(0,x\right)=1+3\ln\left(p\right)+\frac{1}{p}+\ln^{2}\left(p\right)\]

\begin{equation}
-\frac{1}{\pi}\int_{0}^{\infty}\frac{y^{-y}}{\left(p+y\right)^{3}}\sin\left(\pi y\right)\Gamma\left(3,x\left(y+p\right)\right)dy\label{m2}\end{equation}
 corresponding to the first and second momentum respectively.

We can conveniently rewrite these last two equations as

\begin{equation}
\Phi\left(1,x\right)\Phi\left(0,x\right)=-1-\ln\left(p\right)-x\left[1-\Phi\left(0,x\right)\right]-\frac{1}{\pi}\int_{0}^{\infty}\frac{y^{-y}}{\left(p+y\right)^{2}}\sin\left(\pi y\right)\exp\left(-xy\right)dy\label{m1short}\end{equation}
 and

\[
\Phi\left(2,x\right)\Phi\left(0,x\right)=1+\left(3+2x\right)\ln\left(p\right)+\frac{1}{p}+\ln^{2}\left(p\right)+2x+x^{2}\left[1-\Phi\left(0,x\right)\right]\]

\begin{equation}
+2x\Phi\left(1,x\right)\Phi\left(0,x\right)-\frac{2}{\pi}\int_{0}^{\infty}\frac{y^{-y}}{\left(p+y\right)^{3}}\sin\left(\pi y\right)\exp\left(-x\left(y+p\right)\right)dy.\label{m2short}\end{equation}

By simplicity we also divide the range of integration of the integrals
present in (\ref{distrib3}),(\ref{m1short}) and (\ref{m2short})
as $\left[0,1\right]\times\left[1,\infty\right]$ such that we can
easily evaluate them in terms of the two auxiliary functions previously
defined. In particular, all the integrals defined in $\left[1,\infty\right]$
are well defined and consequently they can be evaluated at $p=0$
before being computed. The divergent behaviour in the parameter $p$
of the integrals defined in $\left[0,1\right]$ is extracted by integrating
them by parts. Then, the remaining finite parts are also expressed
in terms of the auxiliary functions.

In the case $n=0$ the integral defined in $\left[0,1\right]$ is
well defined and we do not need to integrate it by parts. Consequently,
Eq. (\ref{distrib3}) provides directly the final expression for the
Landau Distribution

\begin{equation}
\Phi\left(0,x\right)=1-M_{0}\left(x\right)-T_{-101}\left(x\right)\label{finaldistrib}\end{equation}
 which for the first and second momenta one has

\begin{eqnarray}
\Phi\left(1,x\right) & = & -\frac{x\left[1-\Phi\left(0,x\right)\right]+M_{1}\left(x\right)}{\Phi\left(0,x\right)}\nonumber \\
 &  & -\frac{T_{-121}\left(x\right)+\left(1+x\right)T_{-111}\left(x\right)+\pi T_{-112}\left(x\right)}{\Phi\left(0,x\right)}\label{finalfirst}\end{eqnarray}
 and

\begin{eqnarray}
\Phi\left(2,x\right) & = & \frac{1+3x+x^{2}\left[1-\Phi\left(0,x\right)\right]+2x\Phi\left(1,x\right)\Phi\left(0,x\right)}{\Phi\left(0,x\right)}\nonumber \\
 &  & -\frac{\exp\left(-x\right)\left[1-\pi^{2}/3\right]-2M_{2}\left(x\right)-I\left(x\right)}{\Phi\left(0,x\right)}\label{finalsecond}\end{eqnarray}
 respectively, where $I\left(x\right)$ is given by the expression

\begin{eqnarray}
I\left(x\right) & = & -\pi^{2}\left[T_{021}\left(x\right)+2\left(1+x\right)T_{011}\left(x\right)-\frac{\pi^{2}}{3}T_{001}\left(x\right)\right]\nonumber \\
 &  & -\pi T_{-112}\left(x\right)+\left(1+x\right)T_{-111}\left(x\right)-\pi T_{-101}\left(x\right)\nonumber \\
 &  & -\pi\left\{ T_{030}\left(x\right)+3\left(1+x\right)T_{020}\left(x\right)-\frac{\pi^{2}}{3}\left(1+x\right)T_{000}\left(x\right)\right.\nonumber \\
 &  & \left.+\left[2\left(1+x\right)^{2}-\frac{\pi^{2}}{3}\right]T_{010}\left(x\right)\right\} .\label{aux3}\end{eqnarray}

\section{Algorithm}

The nonperturbative expressions (\ref{eq:density}), (\ref{eq:distribution}),
(\ref{eq:momentouno}) and (\ref{eq:momentodos}) are used for $x>3$
, and the corresponding perturbative expressions (\ref{densitypert}),
(\ref{finaldistrib}), (\ref{finalfirst}) and (\ref{finalsecond})
for $\left|x\right|\leq3$. These expressions are the kernel of our
algorithm.

In order to compute the nonperturbative expressions we have provided
the algorithm with the following numerical coefficients

\begin{equation}
B_{kr}=\frac{\left(-\right)^{k+r}A_{kr}}{r!\left(k-r\right)!}\label{eq:bkr}\end{equation}
 where $A_{kr}$ is given by (\ref{eq:akr}). These coefficients were
computed for $k>r$ by the following relationship

\begin{equation}
A_{kr}=\sum_{p=0}^{k-r-1}\left(\begin{array}{c}
k-r-1\\
p\end{array}\right)\left.\frac{\partial^{p}}{\partial a^{p}}\Gamma\left(k+a+1\right)\Psi\left(k-r-1-p,k+a+1\right)\right|_{a=0}\label{eq:akrrec}\end{equation}
 $\Psi\left(n,z\right)$ being the polygamma function\cite{abra}
and $A_{kk}=\Gamma\left(k+1\right)$.

We computed these coefficients through the following numerical recursive
algorithm

\[
for\: j=0,\left(k-r\right)\rightarrow A\left(k,j\right)=\sum_{p=0}^{j-1}\left(\begin{array}{c}
j-1\\
p\end{array}\right)A\left(k,p\right)\Psi\left(j-1-p,k+1\right)\]
 where $A\left(k,0\right)\equiv\Gamma\left(k+1\right)$. Consequently
$A_{kr}=A\left(k,k-r\right)$. Replacing these coefficients into Eq.(\ref{eq:bkr})
we get the numbers $B_{kr}$. This algorithm was worked with MAPLE
software.

Taking into account that our results provide excellent approximations,
considering only a few terms of the corresponding series, we have
just calculated the first 5050 coefficients corresponding to values
of $k=0\cdot\cdot99$ and $r=0\cdot\cdot k$

The computation for the functions $\Phi\left(0,x\right)$, $\Phi\left(1,x\right)$
and $\Phi\left(2,x\right)$ involves, in addition, the evaluation
of the Incomplete Gamma function with complex argument and integer
order $\Gamma\left(n,z\right)$. It is accomplished by using the following
expression\cite{abra}

\[
\Gamma\left(n,z\right)=\left(n-1\right)!\exp\left(-z\right)\sum_{p=0}^{n-1}\frac{z^{p}}{p!}\qquad n\geq1.\]

On the other hand, the perturbative expressions are computed by providing
the algorithm the following numerical coefficients

\begin{equation}
D2_{jnkp}\equiv\frac{D_{jnkp}}{\left(2p+1\right)!4^{p}}\label{Djnkp}\end{equation}
 where $D_{jnkp}$ is given by the Eq.(\ref{coef2}). These coefficients
were numerically computed for $0\leq j\leq3,$ $0\leq n\leq26$, $1\leq k\leq20$
and $0\leq p\leq14$ using MAPLE software.

In addition we have defined auxiliary functions to compute the expressions
(\ref{aux1}), (\ref{aux2}) and (\ref{aux3}).

The data files containing the aforementioned numerical coefficients
(\ref{eq:bkr}) and (\ref{Djnkp}) can be found in the following link:
www.fisica.unlp.edu.ar/download/landau.

\section{Numerical Results}

The test of our algorithm is based on the numerical integration of
the expressions (\ref{eqlandau}) and (\ref{eq:moments}) using MAPLE
software. We found that the algorithm allows to obtain results with
a precision up to thirteen to fourteen significant digits for the
Landau density and distribution and nine to ten for the first and
second momenta. If necessary, higher accuracy can be obtained by adding
more coefficients $B_{kr}$ and $D2_{jnkp}$ into the algorithm.

The set of points chosen to evaluate the aforementioned functions
includes those values presented in Ref. 5 where the precision achieved
certainly depends on the convergence of the expressions obtained.
The corresponding numerical solutions for these functions are tabulated
in the Tables 1,2,3 and 4. We also included in these Tables an estimation
on the order of the series associated with our solutions by which
a lower precision in the numerical results is obtained. Specifically,
we have computed $\left|\Delta f\left(x\right)\right|=\left|f_{s+1}\left(x\right)-f_{s}\left(x\right)\right|\leq10^{-\textrm{d}}$,
where $f\left(x\right)$ represents any of the following four functions;
$\phi\left(x\right)$, $\Phi\left(0,x\right)$, $\Phi\left(1,x\right)$
and $\Phi\left(2,x\right)$, and {}``$s$'' represents the number
of terms needed by the sum over the dummy variable $k$ (Eqs. (\ref{eq:density}),
(\ref{eq:distribution}), (\ref{eq:momentouno}) and (\ref{eq:momentodos}))
or $n$ (Eqs.(\ref{aux1}) and (\ref{aux2})) to obtain a precision
of {}``$d$'' significant digits in the nonperturbative or perturbative
regimes respectively.

Indeed, this information provides insight on how the convergence of
our solutions depend on the point where the function is evaluated
and on the required precision. In particular, we note that faster
evaluations correspond to points localized closer to the origin or
to the positive tails.

For getting some computing time estimation, we developed a partially
optimized program written in Fortran 77 which is based on the above
described algorithm. We found that the typical running time to compute,
with a given accuracy, the four functions over the complete set of
points is, in general, of the order of the fraction of a second in
a common PC processor. For instances, the total number of CPU seconds
that the process spent in user mode is around $0.205$, $0.197$ and
$0.185$ for $d$ equal to $12$, $9$ and $6$ respectively.

\section{Summary and Discussion}

\textit{\emph{The Landau distribution as well as its first and second
momenta are used to describe the energy loss of charged particles
by ionization, particularly within the area of the high energy physics.
At present, just rational approximations and asymptotic expressions
for these functions were available. Numerical algorithms using these
approximate solutions provide a fairly fast computation. In this paper
we presented a}} \textit{\emph{direct calculation of the integral
representation of these functions obtaining perturbative and nonperturvative
solutions expressed in terms of fast convergent series. These solutions
were obtained by using analytical continuation properties and the
expression of the Laplace transform of the Landau density.}} They
have been shown to be very flexible, allowing to combine accuracy
and speed to adapt our approach to the computation of these functions
to many fields of application. We remark that with only few terms
of our series and a fast evaluation of them in all the range of its
variables, we obtain excellent approximations. In fact, we found that
our algorithm \textit{\emph{provides in reasonable computing times
correct results up to 13-14 significant digits on the density and
distribution functions and 9-10 on the first and second momenta.}}
This is clearly a substantial numerical advantage over the existing
programs mentioned above. If necessary, one can improve that precision
by including further coefficients into the algorithm.

Future work involves the obtention of approximate analytical expressions
for the single collision cross section and the electro dynamic properties
of the detector medium\cite{OB}. These quantities, which can be computed
in terms of the Landau distribution, could preserve the main features
of our solutions. In particular they could be similarly expressed
\textit{\emph{in terms of fast convergent series.}} This will surely
improve not only its numerical evaluation but it will also \textit{\emph{allows
to produce suitable fits of the truncated Landau distribution to measured
or simulated energy-loss data.}} In fact, one should obtain an approximate
analytical expression for the dielectric function (or equivalently
the oscillator strength function) which is well-defined over the complete
range of energy. This should remove the presence of non-physical high
energy behaviour introduced in Ref. 12 by using a different approximation
for the Landau distribution.

\section*{Acknowledgments}

We thank Sergio Fanchiotti for stimulating discussion on the algorithm
used. CAGC thanks J. Bernabeu and the Departamento de F\'{\i}sica
Te\'{o}rica, Universidad de Valencia where part of this work was
performed, for the warm hospitality extended to him. CAGC and HF acknowledge
CONICET and Agencia Nacional de Promoci\'{o}n Cient\'{\i}fica for
financial support.

\appendix

\section{Technical Details}

We start by analyzing the expression (\ref{eq:integrals}) $B_{kr}^{n}\left(p,x\right)$
where we cannot, a priori, interchange the operations of limit $p\rightarrow0^{+}$
and the integration. In those cases we are interested in the obtention
of their finite analytic contribution. Their divergent terms cancel
with terms like $p^{p}\left[n\left(n-1\right)\left(n-3/2\right)/p+\left(1+\ln\left(p\right)\right)^{n}\right]$
coming from Eq. (\ref{eq:gralexpression}). Then we solve the other
integrals where the interchange of operations simplifies the calculation.

In the case $n=1$ the divergent integral is just the first one which
is given for $k=0$ and $r=0$, namely

\[
B_{00}^{1}\left(p,x\right)=\begin{array}{l}
\lim\\
p\rightarrow0^{+}\end{array}{\textrm{Im}}\left[\int_{x}^{+\infty}c\exp\left(-pc\right)\frac{1}{\left(c-i\pi\right)}dc\right].\]
 Integrating by parts and expanding for small $p$ we obtain

\begin{equation}
B_{00}^{1}\left(p,x\right)\simeq\pi\left[\gamma+\frac{1}{2}\ln\left(x^{2}+\pi^{2}\right)+\ln\left(p\right)\right]+\mathcal{O}\left(p\right).\label{eq:nonekzero}\end{equation}

The other integrals corresponding to ($k\geq1$) can be rewritten,
after a change of variables, as

\[
B_{kr}^{1}\left(0,x\right)={\textrm{Im}}\sum_{m=0}^{1}\left(\begin{array}{c}
1\\
m\end{array}\right)\left(i\pi\right)^{1-m}\begin{array}{l}
\lim\\
b\rightarrow\infty^{+}\end{array}\left[\int_{\ln\left(x-i\pi\right)}^{\ln\left(b-i\pi\right)}u^{r}\exp\left[-u\left(k-m\right)\right]du\right].\]
 Taking into account that these integrals can be expressed in term
of the incomplete Gamma function, we obtain:

\begin{itemize}
\item for $k>1$
\end{itemize}
\begin{equation}
B_{kr}^{1}\left(0,x\right)={\textrm{Im}}\sum_{m=0}^{1}\left(\begin{array}{c}
1\\
m\end{array}\right)\left(i\pi\right)^{1-m}\frac{1}{\left(k-m\right)^{r+1}}\left[\Gamma\left(r+1,\left(k-m\right)\ln\left(x-i\pi\right)\right)\right]\label{eq:noneklargerone}\end{equation}

\begin{itemize}
\item for $k=1$
\end{itemize}
\begin{equation}
\begin{array}{c}
{\displaystyle B_{1r}^{1}\left(0,x\right)=\pi Re\left[\Gamma\left(r+1,\ln\left(x-i\pi\right)\right)\right]}\\
{\displaystyle +\frac{1}{r+1}\left[1-r+r\ln\left(x^{2}+\pi^{2}\right)\right]\left[\pi/2-\arctan\left(x/\pi\right)\right]}\end{array}\label{eq:nonekone}\end{equation}

Replacing Eqs.(\ref{eq:nonekzero}), (\ref{eq:noneklargerone}) and
(\ref{eq:nonekone}) into Eq.(\ref{eq:q}) we check that the logarithmic
divergent terms finally cancel and the remaining part of the integral
provides the solution given by expression (\ref{eq:momentouno}).

In the case $n=2$ the divergent integrals are for $k=0$ and $k=1$
and they read

\[
B_{00}^{2}\left(p,x\right)=\begin{array}{l}
\lim\\
p\rightarrow0^{+}\end{array}{\textrm{Im}}\left[\int_{x}^{+\infty}c^{2}\exp\left(-pc\right)\frac{1}{\left(c-i\pi\right)}dc\right]\]
 and

\[
B_{1r}^{2}\left(p,x\right)=\begin{array}{l}
\lim\\
p\rightarrow0^{+}\end{array}{\textrm{Im}}\left[\int_{x}^{+\infty}c^{2}\exp\left(-pc\right)\frac{\left[\ln\left(c-i\pi\right)\right]^{r}}{\left(c-i\pi\right)^{2}}dc\right]\]
 respectively. Integrating by parts and expanding for small values
of $p$ we obtain

\begin{itemize}
\item for $k=0$ and $r=0$
\end{itemize}
\begin{equation}
B_{00}^{2}\left(p,x\right)\simeq\pi\left[\frac{1}{p}-x-\frac{\pi}{2}\left[\pi-2\arctan\left(x/\pi\right)\right]\right]+\mathcal{O}\left(p\right)\label{eq:ntwokzero}\end{equation}

\begin{itemize}
\item for $k=1$ and $r=0$
\end{itemize}
\begin{equation}
B_{10}^{2}\left(p,x\right)=\pi\left[\frac{x^{2}}{\left(x^{2}+\pi^{2}\right)}-1-2\left[\gamma+\ln\left(p\right)+\frac{1}{2}\ln\left(x^{2}+\pi^{2}\right)\right]\right]+\mathcal{O}\left(p\right)\label{eq:ntwokonerzero}\end{equation}

\begin{itemize}
\item for $k=1$ and $r=1$
\end{itemize}
\[
B_{11}^{2}\left(p,x\right)=\frac{x^{2}\left[x\left[\arctan\left(x/\pi\right)-\frac{\pi}{2}\right]+\frac{\pi}{2}\ln\left(x^{2}+\pi^{2}\right)+\pi\right]}{x^{2}+\pi^{2}}-2\pi+\gamma\pi+\gamma^{2}\pi+\frac{\pi^{3}}{6}\]

\[
-\frac{\pi}{4}\left[\ln\left(x^{2}+\pi^{2}\right)\right]^{2}-2x\left[\arctan\left(x/\pi\right)-\frac{\pi}{2}\right]+\pi\left[\arctan\left(x/\pi\right)-\frac{\pi}{2}\right]^{2}\]

\begin{equation}
+\pi\left(1+2\gamma\right)\ln\left(p\right)+\pi\left[\ln\left(p\right)\right]^{2}+\mathcal{O}\left(p\right)\label{eq:ntwokonerone}\end{equation}
 The others integrals ($k\geq2$) can be solved in the same way. The
integrals can be written as

\[
B_{kr}^{2}\left(0,x\right)={\textrm{Im}}\sum_{m=0}^{2}\left(\begin{array}{c}
2\\
m\end{array}\right)\left(i\pi\right)^{2-m}\begin{array}{l}
\lim\\
b\rightarrow\infty^{+}\end{array}\left[\int_{\ln\left(x-i\pi\right)}^{\ln\left(b-i\pi\right)}u^{r}\exp\left[-u\left(k-m\right)\right]du\right]\]
 that allows one to obtain

\begin{itemize}
\item for $k>2$
\end{itemize}
\begin{equation}
B_{kr}^{2}\left(0,x\right)={\textrm{Im}}\sum_{m=0}^{2}\left(\begin{array}{c}
2\\
m\end{array}\right)\left(i\pi\right)^{2-m}\frac{1}{\left(k-m\right)^{r+1}}\left[\Gamma\left(r+1,\left(k-m\right)\ln\left(x-i\pi\right)\right)\right]\label{eq:ntwoklargertwo}\end{equation}

\begin{itemize}
\item for $k=2$
\end{itemize}
\begin{equation}
\begin{array}{c}
{\displaystyle B_{2r}^{2}\left(0,x\right)={\textrm{Im}}\left[-\frac{1}{r+1}\left[\left[\ln\left(x-i\pi\right)\right]^{r+1}\right]\right.}\\
{\displaystyle \left.+\sum_{m=0}^{1}\left(\begin{array}{c}
2\\
m\end{array}\right)\left(i\pi\right)^{2-m}\frac{1}{\left(2-m\right)^{r+1}}\left[\Gamma\left(r+1,\left(2-m\right)\ln\left(x-i\pi\right)\right)\right]\right]}\end{array}\label{eq:ntwoktwo}\end{equation}

Replacing now Eqs.(\ref{eq:ntwokzero}), (\ref{eq:ntwokonerzero}),
(\ref{eq:ntwokonerone}), (\ref{eq:ntwoklargertwo}) and (\ref{eq:ntwoktwo})
into Eq.(\ref{eq:q}) we explicitly check that the divergent terms
$1/p$, $3\ln\left(p\right)$, and $\ln\left(p\right)^{2}$ finally
cancel and the rest of the integral provides the solution given by
the expression for the second moment.

\newpage

Table 1

Numerical results for the Landau density. The integers given in columns
$3$-$5$ represent the number ofterms needed to achieve an accuracy
of $12$, $9$ and $6$ significant digits respectively.\newline

\begin{tabular}{|c|c|c|c|c|}
\hline 
$x$&
 $\phi\left(x\right)$&
 $\left|\Delta\phi\right|\leq10^{-12}$&
 $\left|\Delta\phi\right|\leq10^{-9}$&
 $\left|\Delta\phi\right|\leq10^{-6}$\tabularnewline
\hline
$-3$&
 $3.658377588806938E-002$&
 $20$&
 $16$&
 $12$\tabularnewline
\hline
$-2.5$&
 $9.637807619860883E-003$&
 $18$&
 $15$&
 $11$\tabularnewline
\hline
$-2$&
 $4.398547840053790E-002$&
 $17$&
 $14$&
 $10$\tabularnewline
\hline
$-1.5$&
 $0.100550751729020$&
 $15$&
 $12$&
 $9$\tabularnewline
\hline
$0.4$&
 $0.168703915408766$&
 $10$&
 $8$&
 $6$\tabularnewline
\hline
$0.5$&
 $0.165232275482657$&
 $11$&
 $8$&
 $6$\tabularnewline
\hline
$1.5$&
 $0.124221094071471$&
 $16$&
 $13$&
 $9$\tabularnewline
\hline
$3$&
 $7.424765459960213E-002$&
 $21$&
 $18$&
 $14$\tabularnewline
\hline
$4$&
 $5.326865722328474E-002$&
 $29$&
 $17$&
 $8$\tabularnewline
\hline
$5$&
 $3.916341958128983E-002$&
 $21$&
 $13$&
 $7$\tabularnewline
\hline
$8$&
 $1.809028394181073E-002$&
 $10$&
 $9$&
 $7$\tabularnewline
\hline
$10$&
 $1.197648738878854E-002$&
 $11$&
 $9$&
 $6$\tabularnewline
\hline
$15$&
 $5.401336928010399E-003$&
 $10$&
 $8$&
 $6$\tabularnewline
\hline
$20$&
 $3.004979395692253E-003$&
 $10$&
 $8$&
 $6$\tabularnewline
\hline
$30$&
 $1.298671406775338E-003$&
 $9$&
 $7$&
 $5$\tabularnewline
\hline
$60$&
 $3.082734332814474E-004$&
 $8$&
 $6$&
 $4$\tabularnewline
\hline
$100$&
 $1.076112240078372E-004$&
 $7$&
 $6$&
 $4$\tabularnewline
\hline
$500$&
 $4.085712883475095E-006$&
 $5$&
 $4$&
 $2$\tabularnewline
\hline
$1000$&
 $1.012057179716136E-006$&
 $5$&
 $3$&
 $2$\tabularnewline
\hline
$10000$&
 $1.001659320662497E-008$&
 $3$&
 $2$&
 $0$ \tabularnewline
\hline
\end{tabular}

\newpage

Table 2

Numerical results for the Landau distribution.\newline

\begin{tabular}{|c|c|c|c|c|}
\hline 
$x$&
 $\Phi\left(0,x\right)$&
 $\left|\Delta\Phi\right|\leq10^{-12}$&
 $\left|\Delta\Phi\right|\leq10^{-9}$&
 $\left|\Delta\Phi\right|\leq10^{-6}$\tabularnewline
\hline
$-3$&
 $-1.607879630341813E-003$&
 $20$&
 $17$&
 $13$\tabularnewline
\hline
$-2.5$&
 $1.961867909885173E-003$&
 $19$&
 $16$&
 $12$\tabularnewline
\hline
$-2$&
 $1.409435793315372E-002$&
 $17$&
 $14$&
 $11$\tabularnewline
\hline
$-1.5$&
 $4.982427497490971E-002$&
 $15$&
 $13$&
 $9$\tabularnewline
\hline
$0.4$&
 $0.356581474467521$&
 $10$&
 $8$&
 $6$\tabularnewline
\hline
$0.5$&
 $0.373280562479680$&
 $11$&
 $9$&
 $6$\tabularnewline
\hline
$1.5$&
 $0.518343029194760$&
 $16$&
 $13$&
 $10$\tabularnewline
\hline
$3$&
 $0.664205997524218$&
 $21$&
 $18$&
 $14$\tabularnewline
\hline
$4$&
 $0.727271852143877$&
 $26$&
 $15$&
 $8$\tabularnewline
\hline
$5$&
 $0.773026779942834$&
 $18$&
 $12$&
 $7$\tabularnewline
\hline
$8$&
 $0.853459958739293$&
 $12$&
 $9$&
 $7$\tabularnewline
\hline
$10$&
 $0.882938865913564$&
 $11$&
 $9$&
 $7$\tabularnewline
\hline
$15$&
 $0.923293944585224$&
 $11$&
 $9$&
 $6$\tabularnewline
\hline
$20$&
 $0.943462646851257$&
 $10$&
 $8$&
 $6$\tabularnewline
\hline
$30$&
 $0.963250172910313$&
 $10$&
 $8$&
 $6$\tabularnewline
\hline
$60$&
 $0.982280577955328$&
 $9$&
 $7$&
 $5$\tabularnewline
\hline
$100$&
 $0.989571029163290$&
 $8$&
 $6$&
 $5$\tabularnewline
\hline
$500$&
 $0.997976630597665$&
 $6$&
 $5$&
 $3$\tabularnewline
\hline
$1000$&
 $0.998993481974151$&
 $6$&
 $4$&
 $3$\tabularnewline
\hline
$10000$&
 $0.999899912058608$&
 $4$&
 $3$&
 $1$ \tabularnewline
\hline
\end{tabular}

\newpage

Table 3

Numerical results for the first momenta of the density.\newline

\begin{tabular}{|c|c|c|c|}
\hline 
$x$&
 $\Phi\left(1,x\right)$&
 $\left|\Delta\Phi\left(1,x\right)\right|\leq10^{-9}$&
 $\left|\Delta\Phi\left(1,x\right)\right|\leq10^{-6}$\tabularnewline
\hline
$-3$&
 $-2.94418823883569$&
 $21$&
 $18$\tabularnewline
\hline
$-2.5$&
 $-2.67471836189293$&
 $19$&
 $16$\tabularnewline
\hline
$-2$&
 $-2.25828545384356$&
 $17$&
 $13$\tabularnewline
\hline
$-1.5$&
 $-1.86990165100088$&
 $14$&
 $11$\tabularnewline
\hline
$0.4$&
 $-0.679860050176673$&
 $9$&
 $6$\tabularnewline
\hline
$0.5$&
 $-0.629322356963745$&
 $9$&
 $7$\tabularnewline
\hline
$1.5$&
 $-0.180019038237684$&
 $13$&
 $10$\tabularnewline
\hline
$3$&
 $0.339504262821524$&
 $18$&
 $14$\tabularnewline
\hline
$4$&
 $0.611170716736155$&
 $15$&
 $7$\tabularnewline
\hline
$5$&
 $0.839831585464024$&
 $12$&
 $7$\tabularnewline
\hline
$8$&
 $1.35514739591618$&
 $9$&
 $7$\tabularnewline
\hline
$10$&
 $1.60809603494204$&
 $9$&
 $7$\tabularnewline
\hline
$15$&
 $2.06972587987781$&
 $9$&
 $7$\tabularnewline
\hline
$20$&
 $2.39437506213547$&
 $8$&
 $6$\tabularnewline
\hline
$30$&
 $2.84447000017431$&
 $8$&
 $6$\tabularnewline
\hline
$60$&
 $3.59268530830209$&
 $7$&
 $5$\tabularnewline
\hline
$100$&
 $4.12998818980124$&
 $7$&
 $5$\tabularnewline
\hline
$500$&
 $5.77818474253184$&
 $5$&
 $4$\tabularnewline
\hline
$1000$&
 $6.47746787601202$&
 $5$&
 $3$\tabularnewline
\hline
$10000$&
 $8.78657692470135$&
 $4$&
 $3$ \tabularnewline
\hline
\end{tabular}

\newpage

Table 4

Numerical results for the second momenta of the density.\newline

\begin{tabular}{|c|c|c|c|}
\hline 
$x$&
 $\Phi\left(2,x\right)$&
 $\left|\Delta\Phi\left(2,x\right)\right|\leq10^{-9}$&
 $\left|\Delta\Phi\left(2,x\right)\right|\leq10^{-6}$\tabularnewline
\hline
$-3$&
 $8.66852146374132$&
 $24$&
 $21$\tabularnewline
\hline
$-2.5$&
 $7.17765680763846$&
 $22$&
 $19$\tabularnewline
\hline
$-2$&
 $5.14686389834850$&
 $18$&
 $15$\tabularnewline
\hline
$-1.5$&
 $3.58354569903627$&
 $16$&
 $13$\tabularnewline
\hline
$0.4$&
 $0.941718427356982$&
 $9$&
 $7$\tabularnewline
\hline
$0.5$&
 $0.908679035129438$&
 $10$&
 $7$\tabularnewline
\hline
$1.5$&
 $0.944186293458727$&
 $15$&
 $12$\tabularnewline
\hline
$3$&
 $1.82656174148077$&
 $21$&
 $17$\tabularnewline
\hline
$4$&
 $2.72090902630219$&
 $18$&
 $9$\tabularnewline
\hline
$5$&
 $3.74974971267330$&
 $13$&
 $8$\tabularnewline
\hline
$8$&
 $7.21517917106136$&
 $10$&
 $8$\tabularnewline
\hline
$10$&
 $9.64861303605968$&
 $10$&
 $8$\tabularnewline
\hline
$15$&
 $15.7892588337345$&
 $10$&
 $7$\tabularnewline
\hline
$20$&
 $21.8615020920748$&
 $9$&
 $7$\tabularnewline
\hline
$30$&
 $33.7144809006428$&
 $9$&
 $7$\tabularnewline
\hline
$60$&
 $67.7261171683676$&
 $8$&
 $7$\tabularnewline
\hline
$100$&
 $111.315504919124$&
 $8$&
 $6$\tabularnewline
\hline
$500$&
 $525.970930432499$&
 $7$&
 $5$\tabularnewline
\hline
$1000$&
 $1033.83849293122$&
 $6$&
 $5$\tabularnewline
\hline
$10000$&
 $10066.7727522736$&
 $5$&
 $4$ \tabularnewline
\hline
\end{tabular}
\end{document}